\pdfoutput=1

\documentclass[apjl]{emulateapj}

\usepackage{amsmath,amssymb}
\usepackage[colorlinks=true,allcolors=blue,urlcolor=magenta,bookmarks=false,pdfstartview=FitV]{hyperref}
\usepackage{color}

\newcommand{\vK}{v_\mathrm{K}}
\newcommand{\vesc}{v_\mathrm{esc}}
\newcommand{\vej}{v_\mathrm{ej}}
\newcommand{\vhatinf}{\hat{v}_\infty}

\newcommand{\RH}{R_\mathrm{H}}

\newcommand{\symba}{\texttt{SyMBA}}
\newcommand{\swifter}{\texttt{Swifter}}

\begin{document}

\title{Fast litho-panspermia in the habitable zone of the TRAPPIST-1 system}

\author{Sebastiaan Krijt, Timothy J. Bowling, Richard J. Lyons and Fred J. Ciesla}
\email{skrijt@uchicago.edu}

\affil{Department of the Geophysical Sciences, The University of Chicago, 5734 South Ellis Avenue, Chicago, IL 60637, USA}

\begin{abstract} 
With several short-period, Earth-mass planets in the habitable zone, the TRAPPIST-1 system potentially allows litho-panspermia to take place on very short timescales. We investigate the efficiency and speed of inter-planetary material transfer resulting from impacts onto the habitable zone planets. By simulating trajectories of impact ejecta from their moment of ejection until (re-)accretion, we find that transport between the habitable zone planets is fastest for ejection velocities around and just above planetary escape velocity. At these ejection velocities, ${\sim}10\%$ of the ejected material reaches another habitable zone planet within $10^2\mathrm{~yr}$, indicating litho-panspermia can be 4 to 5 orders of magnitude faster in TRAPPIST-1 than in the Solar System.
\end{abstract}

\section{Introduction}\label{sec:intro}
Planet formation is not a 100\% efficient process and most, if not all, mature planetary systems contain a reservoir of comets and asteroids \citep[e.g.,][]{matthews2014}. When these bodies impact planets at high velocities (${\gtrsim}10\mathrm{~km/s}$), material can be ejected at velocities above the planetary escape velocity.  This debris can potentially end up on another planet; such an event explains how the SNC meteorites arrived from Mars \citep[e.g.,][]{gladman1996,worth2013}.   If life developed on the impacted planet, the exchanged material could harbor biological material or organisms that could inoculate the planet onto which it is accreted.  This process is referred to as litho-panspermia \citep{melosh1988,melosh2003}.

In our Solar System, transit times of impact ejecta between terrestrial planets are found to be $10^{6-7}\mathrm{~yr}$ in numerical simulations \citep[e.g.,][]{gladman1996,worth2013}, consistent with cosmic-ray exposure times reported for Martian meteorites found on Earth \citep{eugster2006}. Transit times may be much shorter in other planetary systems, where different orbital architectures result in very different dynamical evolution of ejected debris \citep{steffen2016}. The recently discovered TRAPPIST-1 system is particularly intriguing in this regard, as it consists of seven, nearly Earth-massed planets orbiting an M-type dwarf star. Three of those planets are believed to be within the habitable zone (HZ) and have rocky compositions \citep[e.g.,][]{gillon2016,gillon2017}. With orbital periods on order of days and orbital separations ${<}0.01\mathrm{~AU}$, the transport of material between these planets may be more rapid and efficient than in the Earth-Mars case. \citet{lingam2017} have argued, using an analytical model calibrated to our Solar System, that litho-panspermia could be 2 orders of magnitude faster in TRAPPIST-1 when compared to the Solar System.

In this Letter, we numerically investigate the efficiency and speed of potential litho-panspermia in the TRAPPIST-1 system. Specifically, we seek to quantify what fraction of material that is ejected from one of the HZ planets (e, f, or g) ends up on another planet, how long this material spends in space, and at what speeds it is accreted by another planet.

\begin{deluxetable}{ c  c c c c c c}[t]
\tabletypesize{\scriptsize}
\centering
\tablecaption{Planet properties used in our simulations.}
\tablewidth{0pt}
\tablehead{
\colhead{Planet} & \colhead{$a$}  & \colhead{$R$} & \colhead{$M$} & \colhead{$\RH$} & \colhead{$\vesc$} & \colhead{$\vK$}\\
 & \colhead{$(\mathrm{AU})$}  & \colhead{$(R_\oplus)$} & \colhead{$(M_\oplus)$} & \colhead{$(R_\oplus)$} & \colhead{$(\mathrm{km/s})$} & \colhead{$(\mathrm{km/s})$}
}

\startdata
b& 0.011 & 1.09 & 0.85 & 5.5 & 9.9 & 80 \\
c& 0.015 & 1.06 & 1.38 & 8.8 & 13 & 68 \\
d& 0.021 & 0.77 & 0.41 & 8.4 & 8.2 & 58 \\
e& 0.028 & 0.92 & 0.62 & 13 & 9.2 & 50 \\
f& 0.037 & 1.04 & 0.68 & 17 & 9.0 & 44 \\
g& 0.045 & 1.13 & 1.34 & 26 & 12 & 40 \\
h& 0.063 & 0.76 & 0.40 & 25 & 8.1 & 34
\enddata
\tablecomments{Based on the TRAPPIST-1 system as described in \citet{gillon2017}. The Hill sphere radius is calculated as $\RH=a(M/3M_\star)^{1/3}$, the escape velocity $\vesc=\sqrt{2GM/R}$, and the Keplerian orbital velocity $\vK = \sqrt{GM_\star/a}$.}
\label{tab:setup}
\end{deluxetable}

\section{Simulation setup}\label{sec:setup}
We simulated the trajectories of impact ejecta from their moment of ejection until accretion onto a planet using the {\symba} $N$-body integrator\footnote{Part of the {\swifter} software package, available at \url{http://www.boulder.swri.edu/swifter/}}. This package accounts for the  gravitational interactions between massive bodies (planets and the star) and the gravitational interactions from (all) massive bodies onto test particles (ejecta), but neglects interactions among the test particles and other forces (e.g., radiation pressure, Poynting-Robertson drag). We use a time-step of 0.1 days (1/15th of the orbital period of TRAPPIST-1b), with {\symba} automatically switching to a smaller time-step during close encounters \citep{duncan1998,levison2004}.

The semi-major axes of the orbits of the TRAPPIST-1 planets are listed in Table \ref{tab:setup} along with their radii, masses, and Hill sphere radii. Because the eccentricities and inclinations of the TRAPPIST-1 planets are believed to be very small (\citet{gillon2017} constrain the eccentricities to be ${<}0.085$ and the mutual inclinations to be ${\lesssim}0.2^\circ$), we place the planets on circular and co-planar orbits at the start of our simulations. Over the course of the simulations, interactions between the planets raise the eccentricities of the planetary orbits to $e \lesssim10^{-2}$, consistent with the upper limits given by \citet{gillon2017}. In theory, for eccentric orbits, there will be differences in the trajectories of ejecta released at different times during the planet's orbit \citep[e.g.,][]{jackson2014}. Comparing aphelion and perihelion, the difference in heliocentric distance and orbital velocity $\Delta r/a \approx \Delta v / \vK \approx2e$, so we do not expect significant variations for $e \lesssim10^{-2}$.

The starting positions of the planets (i.e., their initial true anomalies) are chosen at random\footnote{We ran multiple simulations with different starting positions and found no significant differences in the outcomes.}.  At $t=0$, the three planets that are most likely to be in the HZ (e, f, and g) isotropically eject material at a velocity $\vej$ that, in the absence of interactions with other massive bodies, allows them to reach infinity with a remaining velocity $\vhatinf \equiv v_\infty/\vesc$. We will consider ejection velocities\footnote{The ejection velocity normal to the planet's surface can be related through the velocity at infinity by $\vej = (1+\vhatinf)\vesc$.} that result in $0 \leq \vhatinf \leq 2$.

This range covers ejection that is so slow the material barely escapes the planet ($\vhatinf=0$), and high-velocity ejection of material that is launched from the surface at 3 times the planetary escape from the surface ($\vhatinf=2$). While material can be ejected at even higher velocities, we expect the majority of large, mass-dominating, and lightly-shocked ejecta to be just above $\vhatinf=0$ (see Section \ref{sec:implications}). Moreover, at $\vhatinf>2$, the ejection velocity will become comparable to the local Keplerian velocity, causing the ejected material to escape the TRAPPIST-1 system on a hyperbolic orbit \citep{jackson2014}.

To simulate the ejecta, $10^4$ mass-less test particles are released from each planet in the HZ at 0.1 planetary radii above the surface. Ejecta that reaches a heliocentric distance $r<5\times10^{-4}\mathrm{~AU}$ is assumed to be accreted by the primary, and bodies that reach $r>1\mathrm{~AU}$ are taken to be ejected from the system. The simulations are stopped after $10^4\mathrm{~yr}$ (${\sim}300{,}000$ orbits of planet g) because most material has been accreted by then (see Section \ref{sec:results}) and we are focusing on the fastest transport.

While there are considerable uncertainties in the planetary masses as derived from transit timing variations \citep{gillon2017}, it is beyond the scope of this study to address in detail the effects these uncertainties have on the results obtained in Section \ref{sec:results}. However, we are confident our main conclusions are robust because the important quantities scale with planet mass relatively weakly, e.g., $\vesc\propto M^{1/2}$ and $\RH\propto M^{1/3}$.

\begin{figure}[]
\centering
\includegraphics[clip=,width=.95\linewidth]{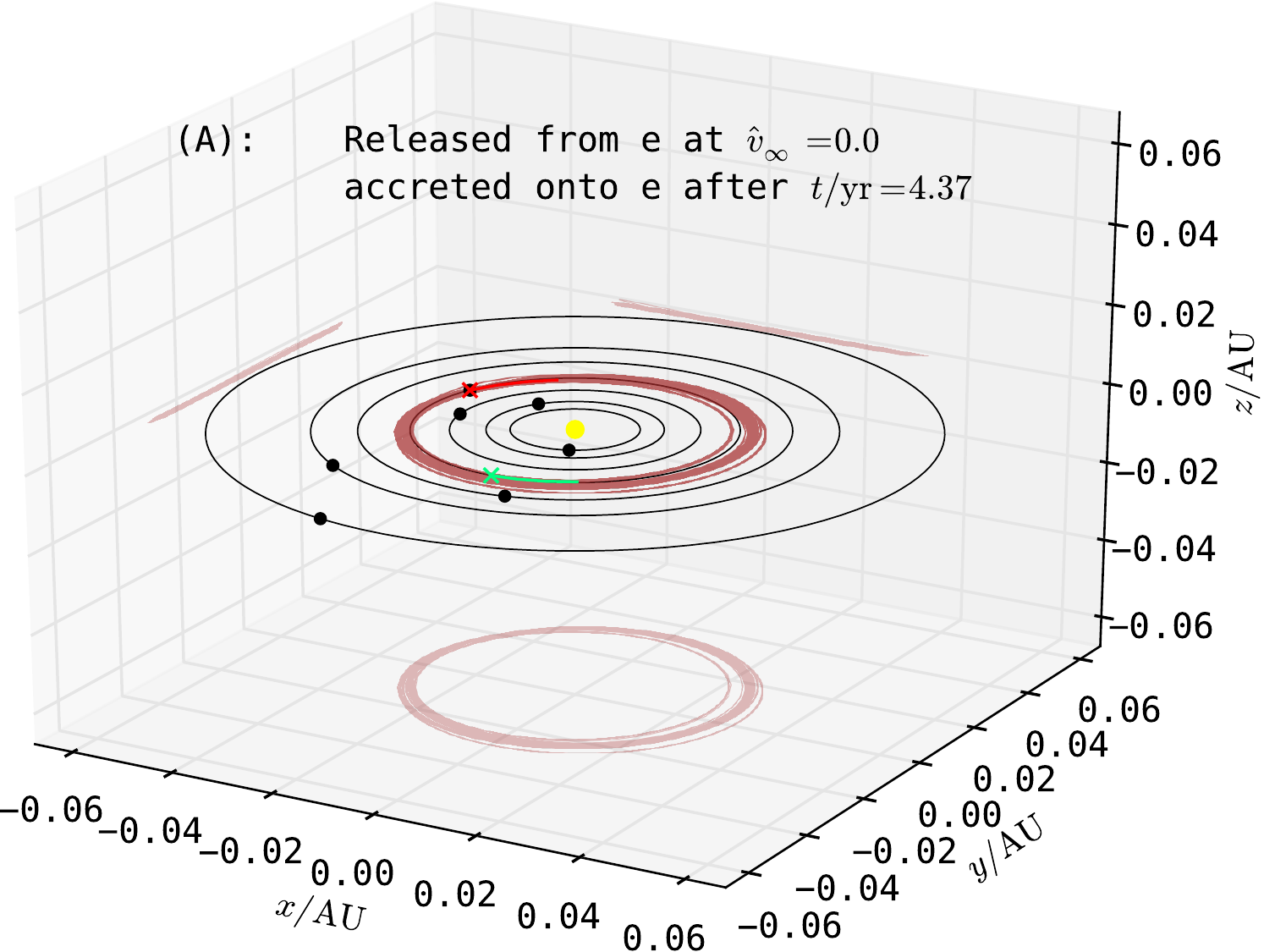}\\
\includegraphics[clip=,width=.95\linewidth]{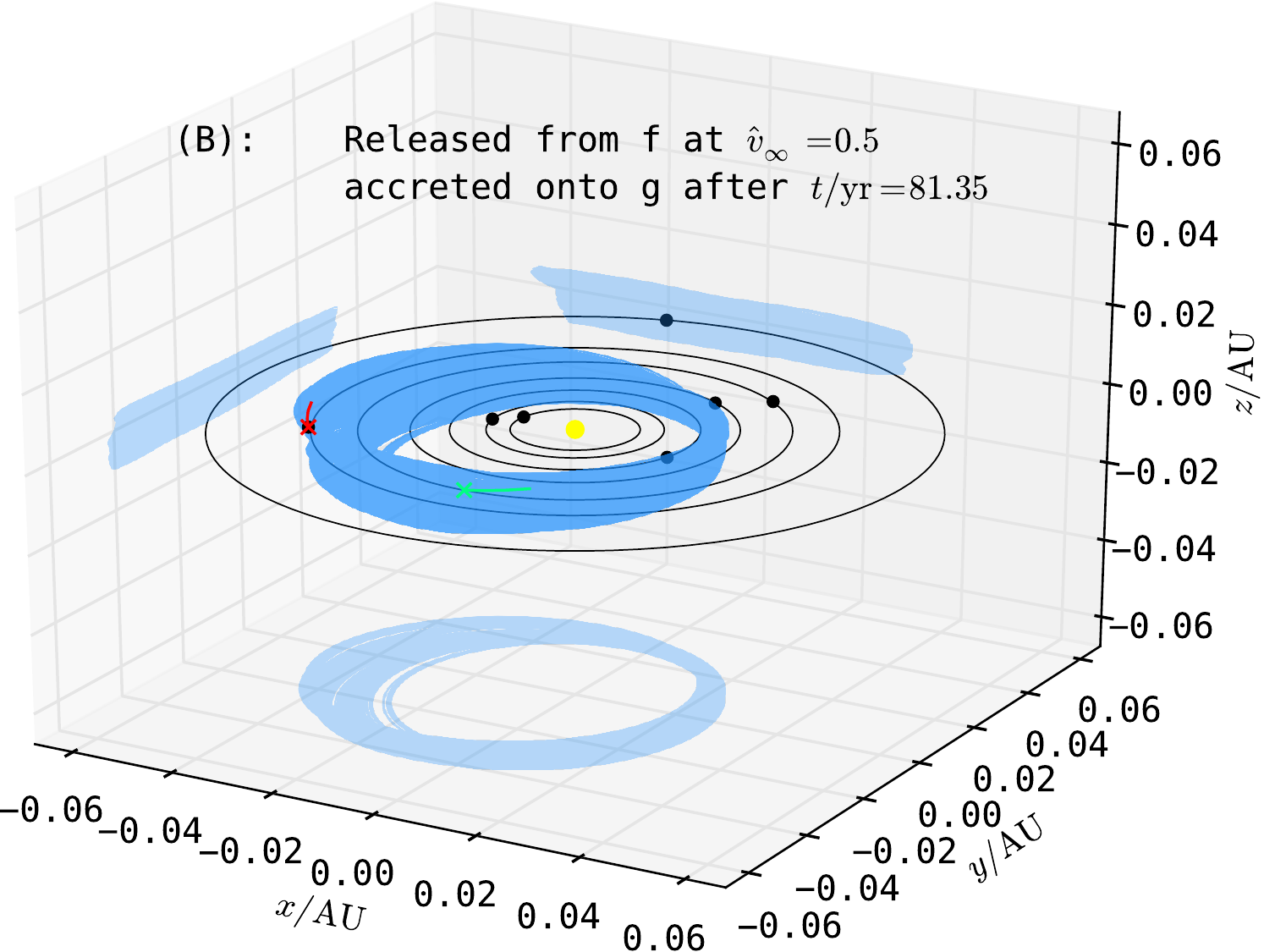}\\
\includegraphics[clip=,width=.95\linewidth]{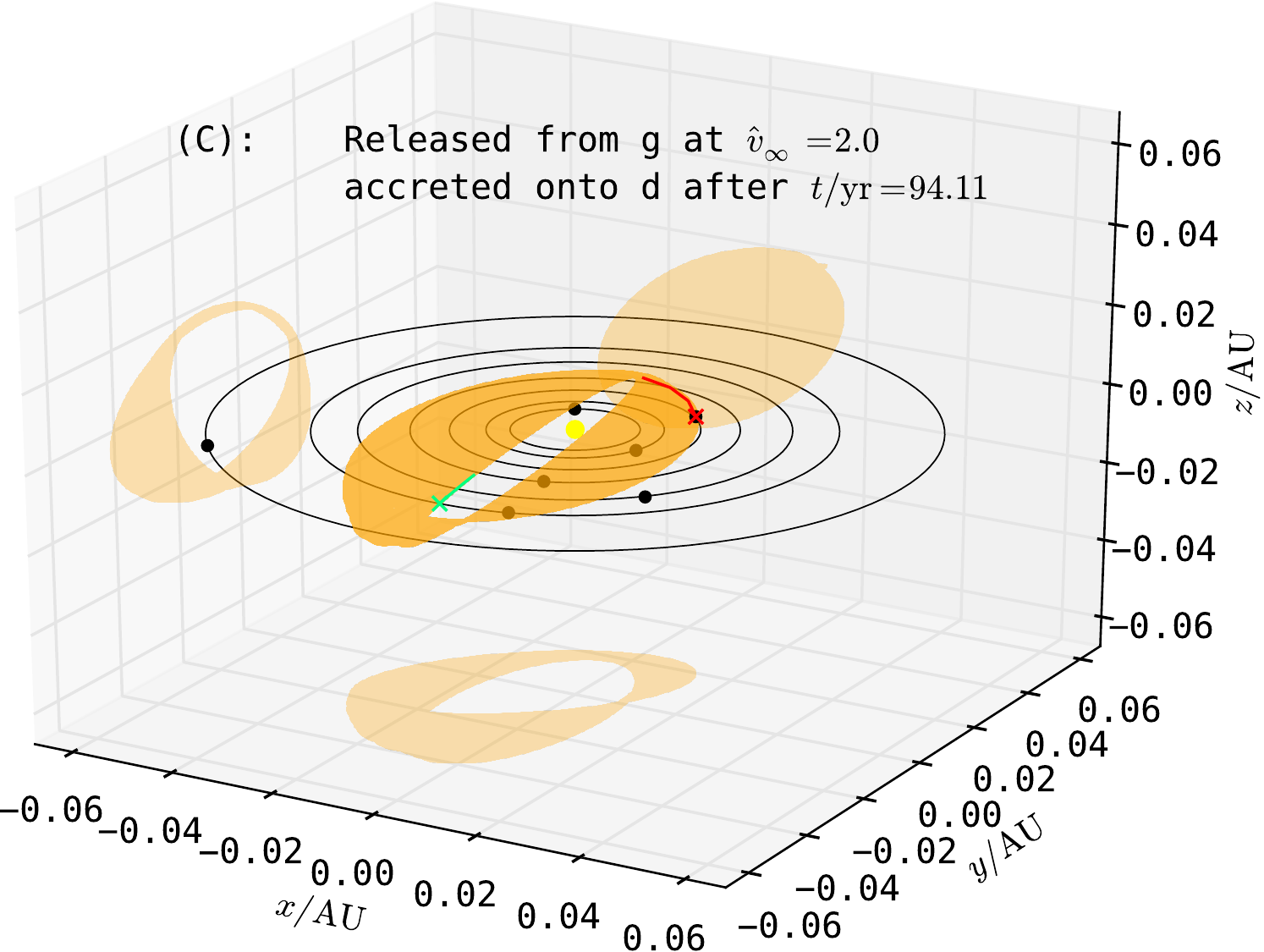}
\caption{Examples of 3D trajectories (and 2D projections) ejecta take after being released from different planets, showing re-accretion onto the same planet (A), outward transport (B) and inward transport (C). Green and red markers show the location and trajectory of ejection and accretion.}
\label{fig:indiv}
\end{figure}

\begin{figure*}[t]
\centering
\includegraphics[clip=,width=.9\linewidth]{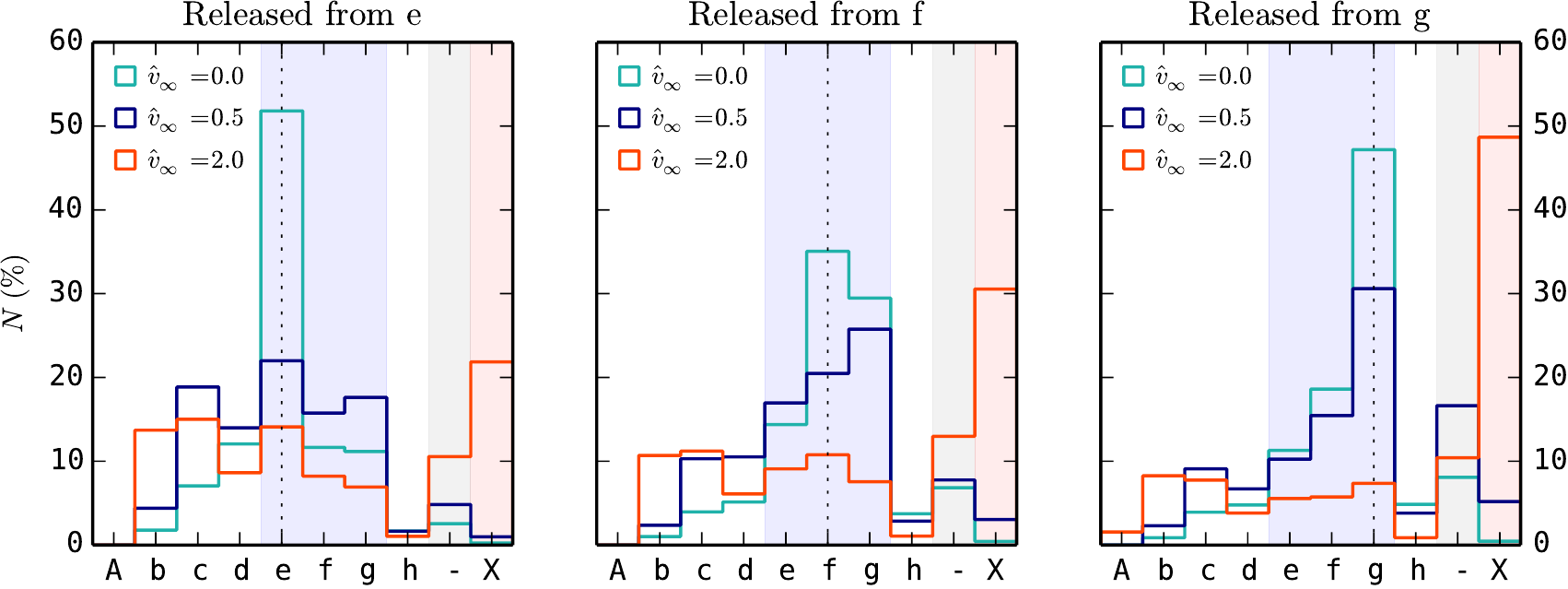}
\caption{Left to right: the fate of ejecta released from planets e, f, and g at different velocities (see text). Letters indicate accretion onto the primary (A) and the seven planets (b through h), X indicates ejection from the TRAPPIST-1 system, and `-' represents ejecta that is still in orbit after the simulated $10^4\mathrm{~yr}$. The blue shaded area corresponds roughly to the HZ.}
\label{fig:fateplot}
\end{figure*}

\section{Results}\label{sec:results}

\subsection{Individual trajectories}
To illustrate the the possible scenarios, Fig. \ref{fig:indiv} shows three trajectories taken by particles ejected from planets e, f, and g at different velocities. In Fig. \ref{fig:indiv}A, material ejected from planet e at a velocity equal to the escape velocity ($\vhatinf=0$) is re-accreted onto the same planet within several years. In Fig. \ref{fig:indiv}B, a particle is ejected at a higher velocity from planet f, and eventually ends up on planet g after ${\sim}80\mathrm{~yr}$. Finally, in Fig. \ref{fig:indiv}C, ejecta released at a high velocity from planet g is eventually accreted by planet d. Generally, higher ejection velocities result in more variation in the eccentricity and inclination of ejecta orbits as the `kick' in velocity these bodies get upon release increases relative to the local Keplerian velocity \citep{jackson2014}.

\subsection{Fate of ejected material}
In Fig. \ref{fig:fateplot}, we plot the outcomes of $10^4$ individual ejecta trajectories for material ejected from planets e, f, and g, at different ejection velocities. Letters indicate (re-)accretion onto the planets and the primary, and the two right-most columns include material that was still in orbit at the end of the simulation or that was ejected from the TRAPPIST-1 system.

Focusing first on relatively slow ejection ($\vhatinf=0$), we find that re-accretion onto the same planet is the dominant outcome. Nonetheless, between ${\sim}45{-}60\%$ of the ejected bodies end up somewhere else, even at these relatively low ejection velocities. In particular, material exchange within the HZ is relatively common (especially between f and g).

At somewhat higher ejection speeds, the distributions in Fig. \ref{fig:fateplot} become broader, with less material being re-accreted by the source planet. In fact, for $\vhatinf=0.5$, the dominant outcome for material released by planet f (middle panel) is accretion by planet g. Accretion onto b, c, and d also becomes more common, as ejecta that is released at higher velocities is more likely to reach highly-eccentric (but bound) orbits that cross the inner parts of the system (see Fig. \ref{fig:indiv}). Qualitatively similar behavior is seen for ejecta originating from Mars and Earth in the Solar System \citep[][Fig. 1]{melosh2003}.

Lastly, for $\vhatinf=2$, ejection from the TRAPPIST-1 system is the dominant outcome, with almost $50\%$ of material released from planet g at these velocities leaving the system on very short timescales. This (virtually instantaneous) ejection from the planetary system occurs when the sum of the ejection velocity and the Keplerian velocity exceeds $\sqrt{2}\vK$, in which case the test particle is released on an unbound orbit \citep{jackson2014}.

\begin{figure*}[t]
\centering
\includegraphics[clip=,width=.95\linewidth]{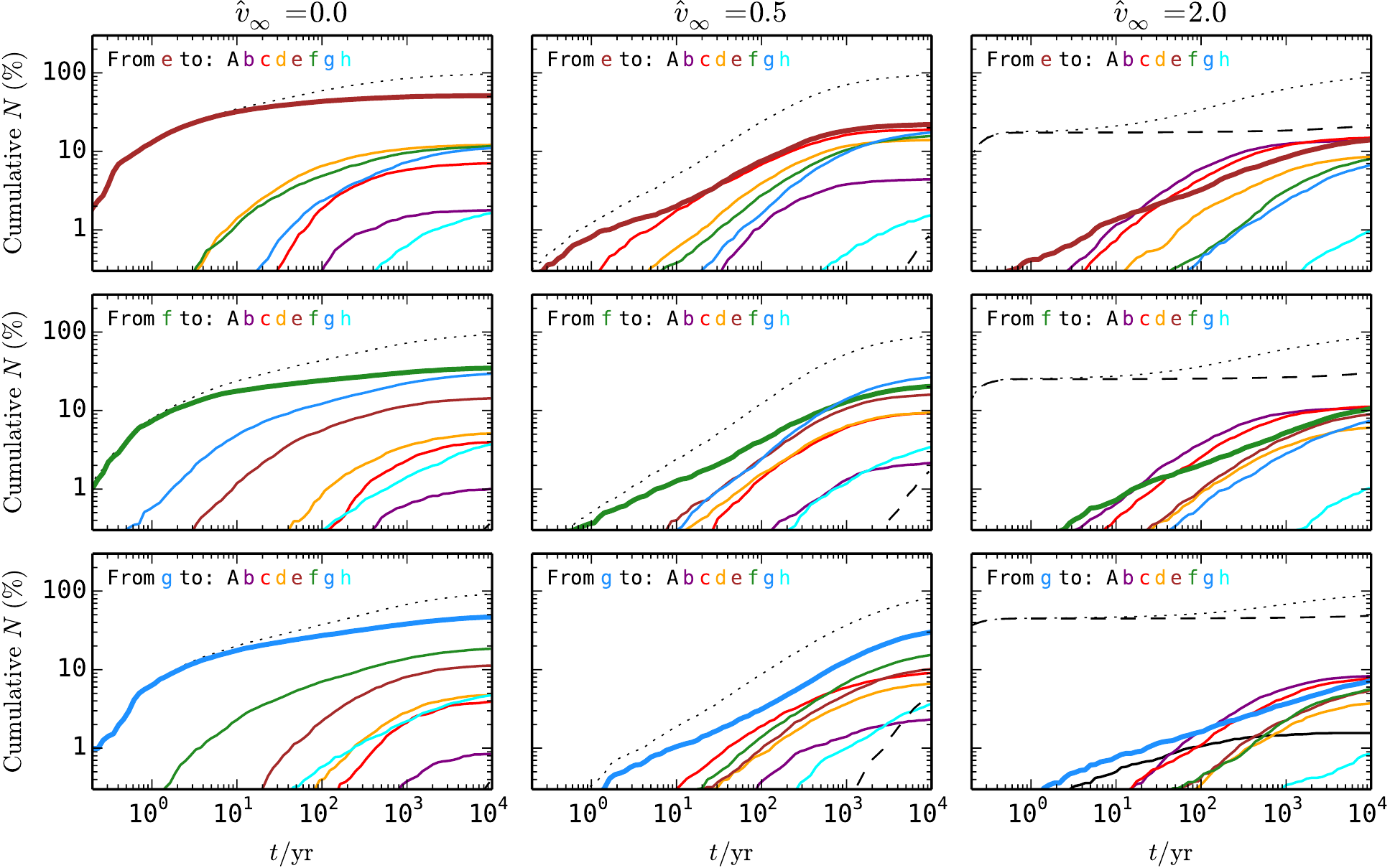}
\caption{Cumulative accretion histories over time. The ejection velocity increases from left to right, and the three rows correspond to ejection by planets e (top), f (middle), and g (bottom). Re-accretion onto the planet of origin is shown by the thickest line, the dashed curve shows complete ejection from the system, and the dotted line shows the sum of all contributions. At the end of the simulations, ${\lesssim}10\%$ of the material released at $t=0$ is still in orbit (see Fig. \ref{fig:fateplot}).} 
\label{fig:cumuplot}
\end{figure*}

\subsection{Time spent `en route'}
In Fig. \ref{fig:cumuplot}, we show the cumulative accretion histories for all bodies in the system for the 9 simulations shown in Fig. \ref{fig:fateplot}. In all low-velocity cases (left column), $10\%$ of the ejected mass returns `home' within ${\sim}1\mathrm{~yr}$. Transport to other planets however is also remarkably fast. Focusing on the HZ planets, the routes\footnote{In this notation f-g should be read as `from f to g'.} f-g, g-f,  f-e, and g-e stand out as being particularly fast for $\vhatinf=0$, transferring between $1{-}10\%$ of ejecta within $100\mathrm{~yr}$. When $\vhatinf$ increased to $0.5$ and $2.0$, both re-accretion onto the planet of origin and transport to other planets become significantly slower and much less material is (re-)accreted in the first $100\mathrm{~yr}$. It is also interesting to look at accretion onto b and c, the inner two planets. In all three low-velocity cases, these two bodies accrete least of the material, and do so relatively late. At the highest ejection velocities however, planets b or c dominate the accretion, even out-pacing re-accretion onto the body of origin.

\subsection{Impact velocities}
\begin{figure}[t]
\centering
\includegraphics[clip=,width=.9\linewidth]{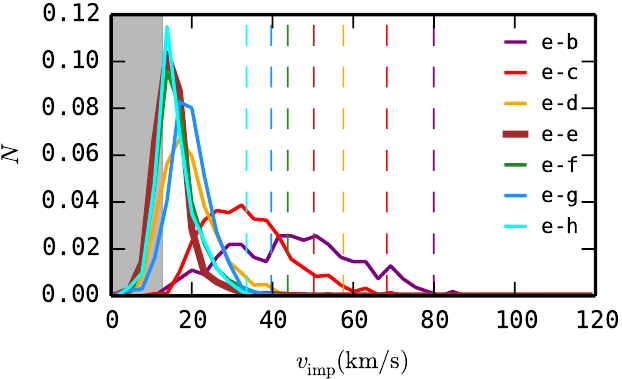}
\caption{Normalized distribution of impact velocities onto different planets for ejecta released from planet e at $\vhatinf=0$. The shaded area indicates the range of planetary escape velocities and the dotted vertical lines represent $\vK$ for planets b through h.}
\label{fig:vimp}
\end{figure}

For organisms to survive re-entry, (relatively) low impact velocities onto the accreting planet are required. Using the final approach of to-be-accreted ejecta (see Fig. \ref{fig:indiv}) we can calculate the (approximate) impact velocity of material. Fig. \ref{fig:vimp} shows the distribution of impact speeds of material ejected from planet e at $\vhatinf=0$ relative to planetary escape and orbital velocities. For transport within the HZ, the most common impact velocities are $10{-}20\mathrm{~km/s}$, within a factor of $2$ of the target escape velocity (Table \ref{tab:setup}). This reflects the fact that most ejecta was released on almost circular and co-planar orbits (Fig. \ref{fig:indiv}A), i.e., the ejection velocity was small compared to the local Keplerian velocity. Reaching b or c requires a more eccentric orbit, and therefore usually results in a considerable contribution of the orbital velocity to $v_\mathrm{imp}$. For increasing $\vhatinf$, the impact velocity distribution shift to higher velocities, with average impact velocities for transport within the HZ reaching $\langle v_\mathrm{imp} \rangle \approx40\mathrm{~km/s}$ for $\vhatinf=2$. Note that the impact velocities shown in Fig. \ref{fig:vimp} correspond to the ejecta's final approach to the targeted planet and do not take in to account potential slowing down of material by the planet's atmosphere.

\section{Implications for (litho-)panspermia}\label{sec:implications}
Comparing the results of Fig. \ref{fig:cumuplot} to typical transfer timescales of $10^{6-7}\mathrm{~yr}$ in the Solar System \citep{gladman1996,worth2013} reveals just how well the proximity of the TRAPPIST-1 planets to each other facilitates material exchange. If the probability of surviving inter-planetary transfer is inversely proportional to the duration of the journey \citep[e.g.,][]{lingam2017}, this indicates the efficiency of litho-panspermia could be $4{-}5$ orders of magnitude more efficient. In addition, the radiation environment (at X-ray and UV wavelengths) around M-stars is expected to be harsh and highly variable \citep{scalo2007}, so short transit times might well be a prerequisite for successful litho-panspermia.

Because the size-/velocity-frequency distribution of the impactor population bombarding the habitable zone of Trappist is unknown, it is impossible to translate our estimates of ejecta flux (Fig. \ref{fig:cumuplot}) into mass flux (kg/yr) of life bearing rock mass. However, using the inner Solar System as a template, we expect the mean impact velocities on TRAPPIST-1 e-g are ${\sim}2$ times the escape velocities of those planets, making it likely that some amount of lightly shocked \citep{melosh1985,johnson2014}, unsterilized \citep[e.g.,][]{mastrapa2001,stoffler2007} material will be transferred between planets every time a large enough impact occurs. Moreover, because both the cumulative amount of mass ejected per impact $M({>} \vej) \propto \vej^{-4}$ \citep{shuvalov2011,johnson2014} and the mean size of ejected fragments $\langle \ell_\mathrm{frag} \rangle \propto \vej^{-2/3}$ \citep{melosh1988} generally decrease with ejection velocity, we expect that material ejected just above escape velocity (i.e., for $\vhatinf$ close to or just above 0) will constitute most of the mass transferred between planets, and will especially dominate the component of transferred mass in large enough fragments that life can survive both irradiation during transfer and heating during re-entry.

\section{Summary}\label{sec:summary}
We have performed $N$-body simulations of the TRAPPIST-1 system to study the fate of impact ejecta released by the three planets that are most likely to be in the habitable zone (e.g., Fig. \ref{fig:indiv}). Our main findings are:

\begin{enumerate}

\item{Material exchange between planets in the TRAPPIST-1 system is very efficient for ejection velocities close to and just above the planetary escape velocity, with $20{-}40\%$ of material ejected at these velocities eventually accreting onto another HZ planet (Fig. \ref{fig:fateplot}).}

\item{Comparing transport timescales between e, f, and g to those of the Mars-Earth route in our own Solar System, we conclude that transferring solids between habitable zone planets in TRAPPIST-1 is up to $4{-}5$ orders of magnitude faster (Fig. \ref{fig:cumuplot}).}

\item{Transport between planets f and g stands out as being particularly fast and effective, with some material being transferred within $10\mathrm{~yr}$ of being released (Fig \ref{fig:cumuplot}).}

\item{The velocities with which transferred material impacts other HZ planets are low ($10{-}20\mathrm{~km/s}$) for material ejected just above escape velocity (Fig. \ref{fig:vimp}); only slightly above the accreting planet's $\vesc$.}

\end{enumerate}

\noindent As we are poised to learn more about the properties of these alien worlds \citep[e.g.,][]{barstow2016,dewit2016}, our results suggest that, as is the case in the Solar System, we should consider the possibility of frequent material exchange between (adjacent) planets in the TRAPPIST-1 system, life-bearing or sterile.

\acknowledgments
The authors thank A. Loeb and M. Lingam for fruitful discussions and the anonymous reviewer for constructive feedback that helped improve the manuscript. This material is based upon work supported by the National Aeronautics and Space Administration under Agreement No. NNX15AD94G for the program ``Earths in Other Solar Systems." The results reported herein benefitted from collaborations and/or information exchange within NASA's Nexus for Exoplanet System Science (NExSS) research coordination network sponsored by NASA's Science Mission Directorate.

\end{document}